\documentclass[12pt,a4paper]{article}
\usepackage{amsfonts}

\newcommand{\cleqn}{\setcounter{equation}{0}}
\newcommand{\clth}{\setcounter{thm}{0}}
\newcommand {\sectionnew}[1]{\section{#1}\cleqn\clth}
\newcommand{\beq}{\begin{equation}}
\newcommand{\eeq}{\end{equation}}
\newcommand{\beqa}{\begin{eqnarray}}
\newcommand{\eeqa}{\end{eqnarray}}
\newcommand{\nn}{\nonumber \\}
\newtheorem{thm}{Theorem}

\newtheorem{prop}[thm]{Proposition}
\newtheorem{lem}[thm]{Lemma}
\newcommand{\proof} {  {\bf{Proof.}}    }
\newcommand {\normprod}[1]{ {\textrm{:}}{#1}{\textrm{:}} } 
\def \W {$W_{1+\infty}$\ }
\def \Wc {$W^{(c)}$\ }
\def \A {\mathcal A}
\def \Beta {\mathrm B}
\def \F {F}  
\def \Ac {${\mathcal{A}}_{c}$\ }
\def \D {{\mathcal D}}
\def \d {{\mathrm d}}
\def \L {{\mathcal L}}
\def \H {{\mathcal H}}
\renewcommand{\vec}{\mathbf}
\def \r {{\vec r}}

\def \Zset {{\mathbb Z}}
\def \Nset {{\mathbb N}}
\begin{document}
\title{\bf A QFT approach to $W_{1+\infty}$ }
\author{
B. N. Bakalov
\thanks{E-mail address:
bbakalov@fmi.uni-sofia.bg }
\\
\normalsize \textit{ Differential Equations,
Department of Mathematics and Informatics, }\\
\normalsize \textit{ Sofia University, Bld. J. Bourchier 5,
BG-1126 Sofia, Bulgaria }
\and
L. S. Georgiev
\thanks{E-mail address:
lgeorg@bgearn.acad.bg }
\\
\normalsize \textit{ Institute for Nuclear Research and Nuclear Energy, }\\
\normalsize \textit{ Bulgarian Academy of Sciences, }\\
\normalsize \textit{ Tsarigradsko Chaussee 72, BG-1784 Sofia, Bulgaria }
\and
I. T. Todorov
\thanks{E-mail address:
todorov@math.mit.edu }
\thanks{ On leave from the Institute for Nuclear
Research and Nuclear Energy, Bulgarian Academy of
Sciences, Tsarigradsko Chaussee 72, BG-1784
Sofia, Bulgaria.}
\\
\normalsize \textit{ Department of Mathematics, Massachusetts
Institute of Technology, } \\
\normalsize \textit{ Cambridge, MA 02139-4307, USA}}
\date{}
\maketitle
\begin{abstract}
\W is defined as an infinite dimensional Lie algebra spanned
by the unit operator and the Laurent modes of a series of local
quasiprimary chiral fields $V^l(z)$ of dimension $l+1$ $(l=0,1,2,\ldots).$
These  fields are neutral with respect  to the u(1) current
$J(z)=V^0(z);$  as a result the $(l+2)$-fold commutator of $J$ with
$V^l$ vanishes. We outline a construction of rational conformal
field theories with stress energy tensor $T(z)=V^1(z)$ whose
chiral algebras include all $V^l$'s. It is pointed out that
earlier work on local extensions of the u(1) current algebra
solves the problem of classifying all such theories for Virasoro central charge
$c=1.$
\end{abstract}
\vspace{-20cm}
\begin{flushright}
{\tt{ hep-th/9512160 }}
\end{flushright}
\vfill
\newpage
{\Large{ {\bf Introduction \\} }}

The infinite dimensional Lie algebra \W has emerged
in various contexts. It was first introduced as the non-trivial
central extension $\widehat{\cal D}$ of the Lie algebra of
differential operators on the circle \cite{kp}. The theory of its
quasifinite highest weight representations was developed by
Kac, Radul et al. \cite{kr,fkrw} (see also \cite{afmo1,afmo2}, the
latter reference containing a review and a bibliography on the subject).
Among the many applications we should like to single those to the
quantum Hall effect \cite{ctz1,ctz2} and to the bispectral problem
(for a recent development and further references see \cite{bhy}).

The present paper starts with a quantum field theoretic (QFT)
characterization of a larger algebra, $\widehat{gl},$ in terms of a
basic bilocal field $V(z_{1},z_{2}).$ \W is spanned by the
local quasiprimary fields $V^l$ entering the expansion of $V$
around  $z_{1}=z_{2}.$ We outline a program for constructing
rational conformal field theory (RCFT) extensions of the chiral
\W algebra and point out that this program has been carried
out in \cite{bmt} and \cite{pt} for Virasoro central charge $c=1$
within the study of local extensions of the u(1) current algebra.

\section{Chiral algebra generated by a bilocal \\ ``dipole
f\/ield''}

We shall begin by describing the QFT counterpart of
$\ \widehat{gl},$ the central extension of the Lie algebra of
infinite matrices with finitely many non-zero
diagonals -- see \cite{fkrw}, Sec.~2. Our starting point will be a
chiral algebra ${\cal A}_{c}={\cal A}_{c}(1,-1)$ generated
by a bilocal field $V(z_{1},z_{2})$ satisfying the commutation
relations (CR)
   \beq \label{1.1}
   \Bigl[V(z_{1},z_{2}),V(z_{3},z_{4})\Bigr]=\Bigl\{V(z_{1},z_{4})+
   \frac{c}{z_{14}}\Bigr\}\delta(z_{23})-\Bigl\{V(z_{3},z_{2})+
   \frac{c}{z_{32}}\Bigr\}\delta(z_{14})
   \eeq
where $z_{ij}=z_{i}-z_{j}$ and the complex
variable $\delta$-function can be defined through its Laurent
expansion (cf. \cite{k})
   \beq \label{1.2a}
   \delta(z_{12})=\frac{1}{z_{12}}+\frac{1}{z_{21}},\quad
   \frac{1}{z_{ij}}=\sum_{n=0}^{\infty} \frac{z_{j}^n}
   {z_{i}^{n+1}}
   \eeq
and is characterized by the properties
   \beq \label{1.2b}
   z_{12}\delta(z_{12})=0,\quad \oint\limits_{|z_1|=|z_2|}
   f(z_{1})\delta(z_{12}) \frac{\d z_{1}}{2\pi{\mathrm i}}=f(z_{2}).
   \eeq
The {\em central charge} $c$ $( [c,V(z_{1},z_{2})]=0 )$
will be regarded as a real number that labels our chiral algebra (along
with the pair $\{1,-1\}$ of opposite charges to be identified
shortly). The inhomogeneous term $$\Psi(z_{1},z_{2};z_{3},z_{4})=\frac{c}
{z_{14}}\delta(z_{23})-\frac{c}{z_{32}}\delta(z_{14})$$ in (\ref{1.1})
satisfies the 2-cocycle condition
     \begin{eqnarray*}
     &&\delta(z_{23})\Psi(z_{1},z_{4};z_{5},z_{6})-
       \delta(z_{14})\Psi(z_{3},z_{2};z_{5},z_{6})+
       \delta(z_{25})\Psi(z_{3},z_{4};z_{1},z_{6})- \nn
     &&\delta(z_{16})\Psi(z_{3},z_{4};z_{5},z_{2})+
       \delta(z_{36})\Psi(z_{5},z_{4};z_{1},z_{2})-
       \delta(z_{45})\Psi(z_{1},z_{2};z_{3},z_{6})=0
     \end{eqnarray*}
which guarantees the Jacobi identity for double commutators. In
order to display the connection of the above Lie algebra
structure with $\widehat{gl}$ it suffices to insert the double
Laurent expansion of $V$
     \beq \label{1.3}
     V(z_{1},z_{2})=\sum_{i,j\in \Zset} E_{ij}z_{1}^{i}z_{2}^{-j-1}
     \eeq
in (\ref{1.1}) and compute the resulting CR for the Weyl matrices
$E_{ij}\colon$
      \beqa\label{1.4}
     \Big[E_{ij},E_{kl}\Bigr]=\delta_{jk}E_{il}-\delta_{il}E_{kj}+c\Bigl
     \{\theta(j)- \theta(i)\Bigr\}\delta_{il}\delta_{jk},\nn
     \theta(i)=1\quad \textrm{for} \quad i>0,\quad \theta(i)+\theta(-i)=1.
      \eeqa
The CR (\ref{1.1}) have a well defined limit when we set, after a
possible differentiation, the arguments of $V$ equal to each
other. The resulting local fields will be viewed as elements of
the same chiral algebra ${\cal A}_{c}.$ The first two of them, the
u(1) current
     \beq \label{1.5}
     J(z)=V(z,z)=\sum_{n\in \Zset} J_{n}z^{-n-1}
     \eeq
and the (chiral) stress energy tensor
     \beq \label{1.6}
     T(z)=\frac{1}{2}\Bigl\{(\partial_{1}-\partial_{2})
     V(z_{1},z_{2})\Bigr\}\Big|_{z_{1}=z_{2}=z}=\sum_{n\in\Zset} L_{n}z^{-n-2}
     \eeq
($\partial_{i}=\frac{\partial}{\partial z_{i}}$) generate a Lie
subalgebra of $\widehat{gl}.$
\newline
Inserting (\ref{1.5}) into (\ref{1.1}) we find
     \beqa \label{1.7a}
     \Bigl[V(z_{1},z_{2}),J(z_{3})\Bigr]&=&\Bigl\{V(z_{1},z_{2})+
     \frac{c}{z_{12}}\Bigr\}\Bigl(\delta(z_{23})-\delta(z_{13})\Bigr)\\
     \Bigl[V(z_{1},z_{2}),J_{n}\Bigr]&=&\Bigl\{V(z_{1},z_{2})+\frac{c}
     {z_{12}}\Bigr\}(z_{2}^n-z_{1}^n).
     \eeqa
These relations justify the $\{1,-1\}$ charge assignment for $V$ (and
the term ``{\em dipole field\/}''). In the limit $z_{2}\rightarrow z_{1}$
we recover the u(1) current algebra:
     \beq \label{1.8}
     \Bigl[J(z_{1}),J(z_{2})\Bigr]=-c\delta^{\prime}(z_{12})
     \Longleftrightarrow \Bigl[J_{m},J_{n}\Bigr]=cm\delta_{m,-n}.
     \eeq
Similarly, (\ref{1.1}) and (\ref{1.6}) imply
     $$
  \Bigl[V(z_1,z_2),T(z_3)\Bigr] = \Bigl\{ \frac{c}{z_{12}^2} -
     \partial_{1} V(z_1,z_2) \Bigr\} \delta(z_{13}) -
     \Bigl\{ \frac{c}{z_{12}^2} +
     \partial_{2} V(z_1,z_2) \Bigr\} \delta(z_{23})
     $$
     \beq\label{1.9a}
  - \; \frac{1}{2}\Bigl\{ V(z_1,z_2) +
     \frac{c}{z_{12}} \Bigr\} \Bigl\{ \delta^{\prime}(z_{13}) +
     \delta^{\prime}(z_{23}) \Bigr\},
     \eeq
     $$
  \Bigl[L_{n},V(z_{1},z_{2})\Bigr] = \Bigl\{z_{1}^n\Bigl(D_{1}
     +\frac{n+1}{2}\Bigr) + z_{2}^n\Bigl(D_{2}+\frac{n+1}{2}\Bigr)
     \Bigr\}V(z_{1},z_{2})
     $$
     \beq\label{1.9b}
  + \; cd_{n}(z_{1},z_{2}),\qquad\qquad D_{i}=z_{i}\partial_{i};
     \eeq
the cocycle $d_{n}(z_{1},z_{2})\;\;(=-d_{n}(z_{2},z_{1}))$ is
homogeneous of degree $n-1$ in its arguments:
     \beqa \label{1.10a}
     &&d_{n}(z_{1},z_{2})=\frac{1}{2z_{12}}\sum_{i=1}^{n-1} (z_{1}^i-z_{2}^i)
     (z_{1}^{n-i}-z_{2}^{n-i})\quad \textrm{for}\;n\geq 2,\quad \\
     &&d_{-n}(z_1,z_2) =
         \frac{1}{z_{1}z_2} d_n \Bigl( \frac{1}{z_1},\frac{1}{z_2} \Bigr)
     \;\; (n\geq 2),\;\; d_n=0 \;\textrm{for}\;n=0,\pm 1.
     \eeqa
The $L_n$ generate the Virasoro algebra:
    \beq \label{1.11}
    \Bigl[L_n,L_m\Bigr]=(n-m)L_{n+m} + c \frac{n^3-n}{12} \delta_{n,-m}.
    \eeq
The vanishing of the cocycle $d_n$ for $n=0,\pm 1$  can be
interpreted by saying that $V$ is a {\em { quasiprimary bilocal field}}.
We recall that $V^l(z)$ is a local quasiprimary field of
(conformal) dimension $l+1$ provided
    \beq \label{1.12}
    \Bigl[L_n,V^l(z)\Bigr]=z^n\Bigl(D+(n+1)(l+1)\Bigr)V^l(z)\quad
    \textrm{for} \quad n=0,\pm 1.
    \eeq
($V^l$ is called {\em{primary\/}} if (\ref{1.12}) holds for all
$n\in\Zset.)$

The importance of the concept of a quasiprimary field stems from
the fact that we shall be dealing, to start with, with the {\em{
vacuum representation }} of the chiral field algebra in which the
lowest weight ({\em {vacuum}}) {\em {vector }} $|0\rangle,$
defined by
    \beqa \label{1.13}
    &&E_{ij}|0\rangle =0 \quad\textrm{for}\quad i\leq j \hfill \nn
    &&\Bigl(\Rightarrow\quad  J_n|0\rangle =0 \;\;\textrm{for}\;\; n\geq 0,
    \qquad L_n|0\rangle =0 \;\;\textrm{for}\;\; n\geq -1\Bigr)
    \eeqa
is {\em M\"{o}bius} or su(1,1) {\em invariant\/}  (it is annihilated
by the {\em{ conformal energy operator\/}} $L_0$ and by $L_{\pm 1}).$
M\"{o}bius invariance determines local (2- and) 3-point functions
of quasiprimary fields (up to a normalization) -- see e.g. \cite{fst}.
Equations (\ref{1.1}), (\ref{1.3}) and (\ref{1.13}) allow to write down
the correlation functions of the basic bilocal field $V.$ We have,
in particular,
     \beqa
        &&\langle 0|V(z_1,z_2)V(z_3,z_4)|0\rangle =
         \frac{c}{z_{14}z_{23}}
\label{1.14a} \hfill \\
         &&\langle 0|V(z_1,z_2)V(z_3,z_4)V(z_5,z_6)|0\rangle =
         \frac{c}{z_{16}z_{23}z_{45}}-\frac{c}{z_{14}z_{25}z_{36}}.
\label{1.14b}
      \eeqa

A realization of the above algebra for a positive integer $c$ is
provided by the composite field (cf. \cite{bprss})
     \beq \label{1.15}
         V_{\psi^*\psi}(z_1,z_2) =
        \sum_{i=1}^c \normprod{ \psi_i^*(z_1)\psi_i(z_2) }
      \eeq
where $\psi_i$ and their conjugates are free Fermi fields
satisfying the canonical anticommutation relations
     \beqa \label{1.16a}
     &&\Bigl[\psi_i(z_1),\psi_j(z_2)\Bigr]_{+}=0=\Bigl[\psi_i^*(z_1),
     \psi_j^*(z_2)\Bigr]_{+}\nn &&\Bigl[\psi_i(z_1),\psi_j^*(z_2)\Bigr]_{+}=
     \delta_{ij}\delta(z_{12})
     \eeqa
or, in Fourier modes,
     \beq\label{1.16b}
     \Bigl[\psi_{i\mu},\psi_{j\nu}^*\Bigr]_{+}=\delta_{\mu,-\nu}\delta_{ij}.
     \eeq
In the vacuum (Neveu--Schwarz) sector $\psi^{(*)}(z)$ is expanded in integer
powers of $z\colon$
     $$\psi^{(*)}(z) = \sum_{n\in\Zset} \psi_{-n-\frac{1}{2}}^{(*)} z^n.$$
The {\em {normal product}} ::  in (\ref{1.15}) is defined by either
subtracting the vacuum expectation value,
     $$\normprod{ \psi^*(z_1)\psi(z_2) } = \psi^*(z_1)\psi(z_2)-
       \frac{1}{z_{12}},$$
or, equivalently, by ordering the Fourier modes:
     \[ \normprod{ \psi_{\mu}^* \psi_{\nu} } =
           \left\{\begin{array}{ll}
                    \quad\psi_{\mu}^*\psi_{\nu} & \mbox{
                           $\textrm{for}\quad \nu > 0$}\\
                    \;-\psi_{\nu}\psi_{\mu}^* & \mbox{
                           $\textrm{for}\quad \nu < 0.$}
            \end{array}\right.\]
\newline
The main goal of this section is to establish an expansion formula for
$V(z_1,z_2)$ in terms of a basis $V^l(z)$ of local quasiprimary fields
(of conformal dimension $l+1)$ that is used in \cite{ctz1,ctz2} for
characterizing the \W algebra.
     \begin{thm}\label{t1.1}
Under the above assumptions the bilocal field $V$ admits the
following expansion in local quasiprimary fields
     \beq\label{1.17}
     V(z_1,z_2)=\sum_{l=0}^{\infty} \frac{(2l+1)!}{(l!)^3}
\int\limits_{z_2}^{z_1} \frac{(z_1-z)^l(z-z_2)^l}{z_{12}^{l+1}} V^l(z) \d z
     \eeq
where $V^l$ satisfy {\rm (\ref{1.12})} and the orthonormalization condition
     \beq\label{1.18}
     \langle0|V^l(z_1)V^{l^{\prime}}(z_2)|0\rangle = c\frac{(l!)^4}{(2l)!}
     z_{12}^{-2l-2}\delta_{ll^{\prime}}.
     \eeq
As a consequence of {\rm (\ref{1.18})} the 3-point function of $V$ and
$V^l$ is computed to be
     \beq\label{1.19}
     \langle0|V(z_1,z_2)V^l(z_3)|0\rangle =l!
     {2l \choose l}^{-1}
     \frac{cz_{12}^l}{(z_{13}z_{23})^{l+1} }.
     \eeq
Conversely, $V^l$ is computed in terms of $V\colon$
     \beq\label{1.20a}
     V^l(z)={2l \choose l}^{-1}
     \lim_{z_1,z_2 \rightarrow z} \sum_{k=0}^l{l \choose k}^{2}
     \partial_1^k(-\partial_2)^{l-k}V(z_1,z_2)
     \eeq
     \beq\label{1.20b}
     =\frac{l!}{(2l)!}\lim_{z_1,z_2 \rightarrow z}
     \partial_1^l(-\partial_2)^{l}\Bigl\{z_{12}^lV(z_1,z_2)\Bigr\}.
     \eeq
Defining the Fourier modes of $V^l$ by the expansion formula
     \beq\label{1.21}
     V^l(z)=\sum_{n\in \Zset} V_n^lz^{-n-l-1}
     \eeq
one deduces that $V_n^l$ satisfy CR of the form
     \beqa\label{1.22}
     \Bigl[V_m^k,V_n^l\Bigr] = (lm-kn) V_{m+n}^{l+k-1} +
     \sum_{\nu =1}^{[\frac{l+k-1}{2}]} p_{\nu}(k,m,l,n) V_{m+n}^{l+k-1-2\nu}
     \nn
      + \; c \frac{(l!)^4}{(2l)!} {l+m \choose m-l-1}
     \delta_{m,-n}\delta_{kl}
     \eeqa
where $p_{\nu}$ are polynomials in their arguments.
      \end{thm}
\proof
We first check the consistency among (\ref{1.17}), (\ref{1.18})
and (\ref{1.19}). To this end we take the matrix element of both
sides of (\ref{1.17}) between $V^l(z_3)|0\rangle$  and $\langle0|$
using (\ref{1.18}) and the integral formula
     \beq\label{1.23}
     \frac{(2l+1)!}{(l!)^2} \int\limits_{z_2}^{z_1} \frac{(z_1-z)^l(z-z_2)^l}
     {z_{12}^{l+1}(z-z_3)^{2l+2}} \d z = \frac{z_{12}^l}
     { (z_{13}z_{23})^{l+1} }.
     \eeq
This expression coincides with the unique up to a normalization M\"{o}bius
invariant 3-point function. Conversely, using the uniqueness of the solution of
the corresponding momenta
problem we deduce the expression for the kernel multiplying $V^l(z)$ in the
integrand of (\ref{1.17}). The numerical coefficients are then computed from
the requirement that the expansion (\ref{1.17}) with $V^l$ satisfying
(\ref{1.19}) reproduces the 4-point function (\ref{1.14a}). The latter property
is a consequence of the identity
     \beq\label{1.24}
     \frac{1}{1-\eta} = 1 + \sum\limits_{l=1}^{\infty} {2l-2 \choose l-1}^{-1}
     \eta^l \F(l,l;2l;\eta)
     \eeq
where the Gauss hypergeometric function is identified from the integral
representation
     \beqa\label{1.25}
      &&\F(l,l;2l;\eta)=\frac{ (2l-1)! }{ [(l-1)!]^2 }
      \frac{ (z_{13}z_{24})^l }{ z_{34}^{2l-1} }
      \int\limits_{z_4}^{z_3} \frac{ (z_3-z)^{l-1} (z-z_4)^{l-1} }
                                   {(z_1-z)^l (z_2-z)^l } \d z  \nonumber \\
      &&=\frac{1}{\Beta(l,l)} \int\limits_{0}^{1}
           \frac{ t^{l-1} (1-t)^{l-1} }{ (1-\eta t)^l } \d t,
      \qquad \Biggl( \Beta(x,y)=\frac{ \Gamma(x)\Gamma(y) }{ \Gamma(x+y) }
              \Biggr),
       \eeqa
$t$ and $\eta$ being the cross-ratios
     \beq\label{1.26}
     t = \frac{ (z-z_4)z_{13} }{ (z_1-z)z_{34} }, \quad
     \eta = \frac{ z_{12}z_{34} }{ z_{13}z_{24} }.
     \eeq
The expression (\ref{1.20a}) for $V^l$ (apart from the normalization factor)
can
be computed in two ways yielding the same answer. First, it follows from $L_1$
covariance (being indeed a special case of eqs. (3.68--71) of \cite{fst}).
Indeed, setting
     \beq\label{1.27}
     V^l(z) = \lim\limits_{z_{1,2}\to z} \D_l(\partial_1 ,\partial_2)
              V(z_1,z_2),
     \eeq
where $\D_l(\alpha,\beta)$ is an yet unknown homogeneous polynomial of degree
$l$ of its arguments, we find from (\ref{1.9b})
     $$
     \Bigl[ L_1, V^l(z) \Bigr] = \lim\limits_{z_{1,2}\to z}
     \D_l(\partial_1,\partial_2)
     \Bigl\{ z_1( z_1\partial_1 + 1) + z_2( z_2\partial_2 + 1) \Bigr\}
     V(z_1,z_2)
     $$
which agrees with (\ref{1.12}) for $n=1$ iff $\D_l$ satisfies the differential
equation
     \beq\label{1.28a}
     \Bigl( \alpha\frac{ \partial^2 }{ \partial\alpha^2 } +
             \beta\frac{ \partial^2 }{ \partial\beta^2 } +
             \frac{ \partial }{ \partial\alpha } +
             \frac{ \partial }{ \partial\beta }    \Bigr)
     \D_l(\alpha,\beta) = 0
     \eeq
which together with (homogenity and) the normalization condition
     \beq\label{1.28b}
     \D_l(1,-1) = 1
     \eeq
yields the operator applied to $V(z_1,z_2)$ in the right hand side of
(\ref{1.20a}). Another way to derive (\ref{1.20a}) is to use the equivalence of
(\ref{1.20a}) and (\ref{1.20b}), to apply the operator
$(\partial_1^n (-\partial_2)^n z_{12}^n)$
to both sides of (\ref{1.17}) and to prove that only the term with $l=n$ does
not vanish in the limit $z_1=z_2.$

We shall use a roundabout road to compute the coefficient to the first term in
the right hand side of (\ref{1.22}) introducing on the way the {\em Taylor
expansion\/} of $V(z_1,z_2)$ around the second point
     \beq\label{1.29}
     V(z_1,z_2) = \sum\limits_{l=0}^{\infty} \frac{ z_{12}^l }{l!} J^l(z_2),
     \quad
     J^l(z) = \partial_1^l V(z_1,z_2) \Big|_{z_1=z_2=z}.
     \eeq
   \begin{lem}\label{lem1.2}
The fields $J^l$ satisfy the local CR
     \beqa\label{1.30a}
     \Bigl[J^k(z_1),J^l(z_2)\Bigr] &=&\sum\limits_{\nu\geq 1}
     \Bigl\{ (-1)^{\nu} {l \choose \nu} J^{k+l-\nu}(z_2) -
     { k \choose \nu } J^{k+l-\nu}(z_1)  \Bigr\} \delta^{(\nu)}(z_{12})\nn
       &+&c (-1)^{l+1} \frac{ k! l! }{ (k+l+1)! } \delta^{(k+l+1)}(z_{12})
\hfill  \eeqa
or, in Fourier--Laurent modes, for
     $$J^l(z) = \sum_{n\in \Zset} J^l_n z^{-l-n-1},$$
     \beqa\label{1.30b}
     \Bigl[J^k_m , J^l_n\Bigr]&=& (lm-kn) J^{l+k-1}_{m+n}  \hfill \nn
     &&+ \; \Bigl\{ l(l-1)m(m+2k-1) - k(k-1)n(n+2l-1)\Bigr\}
         J^{l+k-2}_{m+n}\hfill\nn
     && + \cdots + (-1)^k c k! l! {k+m \choose m-l-1} \delta_{m,-n}.
     \eeqa
   \end{lem}
The {\bf proof of the lemma} consists in a straightforward computation
starting from (\ref{1.1}) and (\ref{1.29}) and using identities of the type
     \beq\label{1.31}
     \Bigl[\partial_2^l\delta(z_{12})\Bigr]\frac{z_{12}^{l-n}}{(l-n)!}
     =(-1)^n
     {l \choose n}\delta^{(n)}(z_{12}),
     \quad \delta^{(m)}(z_{12}) z_{12}^n = 0 \quad \textrm{for} \; n>m.
     \eeq
     \begin{lem}\label{lem1.3}
The quasiprimary fields $V^l$ are expressed in terms of $J^l$ and their
derivatives by
    \beq\label{1.32}
    V^l(z)={2l \choose l}^{-1}
    \sum_{k=0}^l (-1)^k{ l \choose k }{ 2l-k  \choose l }
    \partial^kJ^{l-k}(z).
    \eeq
    \end{lem}
To {\bf prove} (\ref{1.32}) we express $-\partial_2$ in (\ref{1.20b})
as $\partial_1-(\partial_1 + \partial_2).$

The term $(lm-kn)V^{l+k-1}_{m+n}$ in the right hand side of (\ref{1.22})
is obtained from the corresponding term in the commutator of $J$ modes
(\ref{1.30b}) by noting that the coefficient to $J^l$ in the
expansion (\ref{1.32}) is $1.$ This completes the proof of Theorem
\ref{t1.1}.

There are two obvious advantages in computing correlation functions in
terms of quasiprimary fields: the orthogonality relation (\ref{1.18})
(which has no match for $J^l$ fields) and the possibility to use M\"{o}bius
invariance which determines the 3-point functions. The spectrum of the
Cartan subalgebra is, however, easier to write down in terms of
representatives of the differential operators $D^l$ (see \cite{kr} and
Sec.~2 below).

\sectionnew{ The subalgebra \Wc of \Ac and its QFT representations }

Let \Wc be the associative subalgebra of \Ac spanned by (the vacuum
representation of) finite linear combinations of the quasiprimary fields $V^l.$
It can be viewed as a covering of the (vacuum representation of)
$W_{1+\infty}.$ The CR (\ref{1.22}) have an important corollary which
characterizes $W^{(c)}.$
     \begin{prop}\label{prop2.1}
The $(l+1)$-fold commutator of current components $J_{m} = V^{0}_{m}$ with
$V^{l}(z)$ is a $c$-number:
    \beq \label{2.1}
    \Bigl[ J_{m_0}, \Bigl[ J_{m_1},\ldots \Bigl[ J_{m_l},V^{l}(z) \Bigr] \ldots
    \Bigr]\Bigr] = c l! \prod\limits_{\nu = 0}^{l} (m_{\nu}
                   z^{m_{\nu}-1}).
     \eeq
    \end{prop}

It follows that {\em the derivative (commutator) action of the Heisenberg
algebra generated by $J_{m}$ is nilpotent\/} on finite linear combinations of
$V^{l}.$ This motivates the following {\bf{definition}}.
{\em An element $A\in$\Ac is said to belong to \Wc iff there exists a positive
integer $N=N(A)$ such that for any choice of the indices $m_{1},\ldots, m_{N}
\in \Zset$  the $N$-fold commutator $[J_{m_1},\ldots [J_{m_N},A] \ldots ]$
vanishes.}

It follows that the associative local chiral algebra corresponding to \W is
generated by the quasiprimary fields $V^{l}.$
We proceed to summarizing the results of \cite{kr,fkrw} on the
unitary {\em lowest weight modules\/} $\L_c$ of $\widehat\D.$
Denote by $W(z^m f(D))$ the image of the differential operator $z^m f(D)$ in
${\mathrm{End}}\, \L_c.$ We shall also use this notation for
quasipolynomials $f$ (i.e., for polynomials of $D$ and of $e^{\lambda D}).$
The basic CR can be written in a compact form \cite{kr}
    \beqa\label{2.2}
     &&\Bigl[W(z^{m}e^{xD}),W(z^{n}e^{yD})\Bigr] =   \nonumber \\*
     &&(e^{nx}-e^{my})W(z^{m+n}e^{(x+y)D})
      - c \frac{e^{nx}-e^{my}}{e^{x+y}-1} \delta_{m,-n}.
     \eeqa
The positive energy unitary irreducible representations of $\widehat\D =\ $\W
correspond to positive integers $c$ and are labeled by $c$ real charges
$\r = (r_1,\ldots ,r_c).$ Let $|\r\rangle$ be the corresponding minimal energy
state satisfying
     \beq\label{2.3}
     \{ W(-D^l) - \lambda_l \}|\r\rangle = 0 = W(-z^m D^l)|\r\rangle
     \quad \textrm{for} \; m \geq 0.
      \eeq
The eigenvalues $\lambda_l$ of Cartan operators are then recovered from the
generating function \cite{kr}
      \beq\label{2.4}
      \Delta_{\r}(x) := \sum\limits_{l=0}^{\infty} \lambda_l \frac{x^l}{l!} =
      \sum_{i=1}^{c} \frac{e^{r_{i}x}-1}{e^{x}-1}.
      \eeq
We have, in particular,
    \begin{eqnarray*} \lambda_{0} = \sum_{i=1}^{c} r_i ,\;
      \lambda_{1} = \frac{1}{2} \sum_{i=1}^{c} (r_{i}^{2} - r_{i}) ,\;
      \lambda_{2} = \sum_{i=1}^{c} \Bigl(\frac{ r_{i}^{3} }{3} -
                    \frac{ r_{i}^{2} }{2}  + \frac{ r_{i} }{3}\Bigr) .
     \end{eqnarray*}

We should like, to begin with, to reexpress these results in terms of the
quasiprimary fields $V^l$ and their Laurent modes. This is done by using the
modes $J^l_n$ (appearing in (\ref{1.30b})) as intermediary. We have, according
to \cite{kr},
    \beq\label{2.5}
    J^l_n = -W(z^n [D]_l), \quad [D]_0 = 1, [D]_l = D(D-1)\ldots(D-l+1).
    \eeq
Using the relation (\ref{1.32}) between $V^l$ and $\{J^{l-k}\}_{k=0,\ldots,l}$
we find
     \beq\label{2.6a}
     V^l_n = W(z^n f_{ln}(D))
     \eeq
where $f_{ln}$ are polynomials of degree $l\colon$
     \beq\label{2.6b}
     f_{ln}(D) = - {2l \choose l}^{-1} \sum\limits_{\nu=0}^{l} (-1)^{l-\nu}
     {l \choose \nu}^{2} [D]_{\nu} [-D-n-1]_{l-\nu} \;\; (= -D^l + \cdots)
     \eeq
(we have noted that $\sum {l \choose \nu}^{2} = {2l \choose l} ).$

The spectrum of the zero modes $V^l_0$ can be computed independently starting
 from the expectation value of the dipole field $V$ in the ground state
$|\r\rangle\colon$
     \beq\label{2.7}
     \langle\r|V(z,w)|\r\rangle = \frac{1}{z-w} \sum\limits_{i=1}^{c}
     \Bigl\{ {\Bigl(\frac{z}{w}\Bigr)}^{r_i} - 1 \Bigr\} .
     \eeq
Inserting in it the expansion (\ref{1.17}) we find
     \beq\label{2.8}
     \sum\limits_{i=1}^{c} \Bigl\{ {\Bigl(\frac{z}{w}\Bigr)}^{r_i} - 1 \Bigr\}=
     \sum\limits_{l=0}^{\infty} \frac{1}{l!} V_l(\r)
     \Bigl\{ {\Bigl( 1-\frac{w}{z} \Bigr)}^{l+1}
             \F \Bigl( l+1,l+1;2l+2;1-\frac{w}{z} \Bigr) \Bigr\}
     \eeq
($\F$ being the hypergeometric function (\ref{1.25})). The eigenvalues of the
zero modes
     \beq\label{2.9}
     V_l(\r) = \langle\r|V^l_0|\r\rangle  \qquad (\r=(r_1,\ldots,r_c))
     \eeq
are thus computed from the recurrence relation
     \beq\label{2.10a}
     \sum\limits_{i=1}^{c} \frac{r_i(r_i+1)\ldots(r_i+n)} {(n+1)!} =
     \sum\limits_{l=0}^{n} {n \choose l}
     \frac{ n!(2l+1)! }{ {(l!)}^2 (l+n+1)! } V_l(\r).
     \eeq
Another way to compute $V_l(\r)$ is to apply the operator
$(\partial_1^l (-\partial_2)^l z_{12}^l)$
to both sides of (\ref{2.7}) and to use (\ref{1.20b}). The result is
     \beqa
\hspace*{-42pt}
    && V_0(\r) =  \sum r_i \;\; (= \lambda_{0}), \hfill\nonumber \\
\hspace*{-42pt}
    && V_l(\r) = \Beta(l,l+1) \sum\limits_{k=0}^{l-1}
     {l \choose k} {l \choose k+1}
     \sum\limits_{i=1}^{c} r_i \frac{\Gamma(r_i+l-k)}{\Gamma(r_i-k)},
     \quad l \geq 1.      \hfill \label{2.10b}
       \eeqa
We find, in particular,
      \beqa
     &&V_1(\r)  =   \frac{1}{2}\sum r_{i}^{2} \;\; (= \lambda_{1} +
     \frac{1}{2}\lambda_0), \quad
     V_2(\r) =  \frac{1}{3}\sum r_{i}^{3} \;\; (= \lambda_{2} - \lambda_{1}
     - \frac{1}{6}\lambda_0),
\hfill \nonumber \\
     &&V_3(\r)  =   \frac{1}{4}\sum \Bigl( r_{i}^{4} + \frac{1}{5} r_{i}^{2}
\Bigr), \quad
     V_4(\r) =  \sum \Bigl( \frac{1}{5} r_{i}^{5} + \frac{1}{7}
r_{i}^{3} \Bigr) , \ldots  \hfill \nonumber \\
     &&V_l(-\r) =  (-1)^{l+1} V_l(\r). \hfill \label{2.11}
     \eeqa
The first two eigenvalues giving the total charge and the energy of the ground
state justify the use of the term ``charges'' for the representation labels
$r_i.$
For $c \geq 1$ a representation of lowest weight $\r$ is {\em degenerate \/}
provided some of the differences $r_i-r_j$ (for $i \neq j)$ are integers. It is
{\em maximally degenerate \/} if all such differences are integers.
We shall say that a (reducible) positive energy \Wc module $\H_c$ gives rise
to a {\em QFT representation \/} of \W if all its irreducible positive energy
submodules $\L_c (\r)$ are generated from the vacuum sector by
$W^{(c)}$-primary chiral vertex operators $\phi_{\r}(z)$ such that
     \beq\label{2.12}
     \phi_{\r}(z) |0\rangle = e^{zL_{-1}} |\r\rangle
     \eeq
and if the resulting ``field algebra'' is closed under fusion. (For
non-degenerate representations this means that the lowest weights form an
abelian group: if $|\r\rangle,|{\vec s}\rangle \in \H_c$ then also
$|\r+{\vec s}\rangle \in \H_c.$ The fusion rules for degenerate representations
are also written down in \cite{fkrw}.)

\sectionnew{ RCFT extensions of $W^{(c)}.$ The case $c=1.$ }

We set the problem of classifying the local extensions of the chiral algebra
\Wc (with the same central charge $c$) which admit a finite number of QFT
representations $\pi_{\nu}$ such that their specialized characters
     \beq\label{3.1}
     {\mathrm{ch}}_{\nu}(\tau) = {\mathrm{tr}}_{\pi_{\nu}}
      q^{L_0-\frac{c}{24}},
     \quad q = e^{ 2\pi{\mathrm{i}}\tau }, \, {\mathrm{Im}}\tau > 0
     \eeq
span a modular (i.e. ${\mathrm{SL}}(2,\Zset))$ invariant space.

We shall present the solution to this problem for $c=1$ and shall end up with
some remarks concerning the general case. We note that the algebra \Wc with
$c>1$ can be written as the tensor product of a $W^{(1)}$ factor and a chiral
algebra of central charge $c-1$ that involves no U(1) current in the
degenerate case (see \cite{ctz2}, Sec.~3). Thus the $c=1$ theory is a necessary
ingredient for the solution of the general problem.

The Bose--Fermi ($\Zset_2$-graded local) chiral algebras of $c=1$ containing
$W^{(1)}$ coincide with the (local or Fermi local) extensions of the u(1)
current algebra classified in \cite{bmt} and \cite{pt}. Each such extension is
generated by a pair of oppositely charged $\widehat{\mathrm u}(1)$ primary
fields $\psi(z,\pm g)$ with  $g^2 \in\Nset$ satisfying the $\Zset_2$-graded
(Bose--Fermi) locality
     \beq\label{3.2}
     z_{12}^{g^2} \Bigl\{ \psi(z_1,g)\psi(z_2,\pm g) -
     (-1)^{g^2}\psi(z_2,\pm g)\psi(z_1,g)  \Bigr\} = 0.
     \eeq
The condition that $\psi$ is $\widehat{\mathrm u}(1)$ primary means that it
satisfies the operator Ward identities
     \beqa
     &&\Bigl[ J(z_1),\psi(z_2,g) \Bigr] = g\delta(z_{12})\psi(z_2,g)
\label{3.3a} \\
     &&\Bigl[ T(z_1),\psi(z_2,g) \Bigr] = -\Delta_{g}\delta'(z_{12})\psi(z_2,g)
                              + \delta(z_{12})\psi'(z_2,g). \label{3.3b}
      \eeqa
The compatibility of these relations, together with the Sugawara formula
     \beq\label{3.4}
     T(z) = \frac{1}{2} \normprod{ J(z)^2 },
     \eeq
which is a consequence of the \W CR (\ref{1.22}) for $c=1,$ yield the relation
between charge and conformal dimension
     \beq\label{3.5}
     \Delta_g = \frac{1}{2} g^2 \;\; (\in\frac{1}{2}\Nset)
     \eeq
as well as the $\widehat{\mathrm u}(1)$ counterpart of the
Knizhnik--Zamolodchikov equation \cite{t}:
     \beq\label{3.6a}
     \psi'(z,g) = g \normprod{ J(z)\psi(z,g) } = g \Bigl\{ J_{(+)}(z)\psi(z,g)
+                                           \psi(z,g)J_{(-)}(z) \Bigr\}
     \eeq
where
     \beq\label{3.6b}
     J_{(+)}(z) = \sum\limits_{n=1}^{\infty} J_{-n}z^{n-1} = J(z) - J_{(-)}(z).
     \eeq
The QFT representations of the resulting extended algebra $\A (g^2) \supset
W^{(1)}$ which respect the $\Zset_2$-graded locality of $\psi(z,\pm g)$ are
generated by a set of $2g^2$ (multivalued, fractional spin) charge fields
$\phi(z,e_l)$ (including the unit operator) with charges
     \beq\label{3.7}
     e_l = \frac{l}{2g}, \quad 1-g^2 \leq l \leq g^2.
     \eeq
The relation
     \beq\label{3.8}
     \psi(e^{2\pi{\mathrm i}}z,g) = \psi(z,g) e^{2\pi{\mathrm i} g J_0}
     \eeq
implies that $\psi(z,\pm g)$ are single or double valued in the sector $\H_l$
with lowest weight vector $|e_l\rangle$ depending on the parity of $l$ (since
$e^{2\pi{\mathrm i} g e_l} = (-1)^l ).$ If $\psi(z,\pm g)$ are Bose fields
(i.e.
if $g^2$ is even) then $\H_l$ with odd $l$ define $\Zset _2${\em{-twisted
sectors}}. The subset of even $l$'s gives rise to a modular invariant local
theory in that case.
The specialized characters (\ref{3.1}) in the sectors $\H_l$ are given by
\cite{pt}
     \beq\label{3.9}
     K_l(\tau,g^2) = {\mathrm{tr}}_{\H_l} q^{ L_0-\frac{1}{24} } =
     \frac{1}{\eta(\tau)} \sum\limits_{n}  q^{ \frac{1}{2}(ng +
\frac{l}{2g})^2}
     \eeq
where $\eta(\tau)$ is the Dedekind function
     \beq\label{3.10}
     \eta(\tau) = q^{\frac{1}{24} } \prod\limits_{\nu=1}^{\infty} (1-q^{\nu})
     \eeq
and the sum in (\ref{3.9}) runs over all integers $(n\in\Zset).$ They span an
${\mathrm{SL}}(2,\Zset)$ invariant space \cite{pt} (for a review -- see
\cite{fst}, Sec.~7). We find as special cases the extensions by a pair of free
Fermi fields (satisfying (\ref{1.16a})) for $g^2 = 1$, and the level 1 su(2)
current algebra for $g^2 = 2.$ In both cases the resulting RCFT has two
(untwisted) sectors.

If $g^2$ is not square free, $g^2 = k^2 g_0^2$ $(k,g_0^2 \in\Nset)$ then the
chiral algebra can be further extended, $\A (g^2) \subset \A (g_0^2).$
If we define a $\Zset_k$-automorphism group of $\A (g_0^2)$ generated by
     \beq\label{3.11}
     \zeta^{\frac{J_0}{g_0}} \psi(z,\pm g_0) \zeta^{-\frac{J_0}{g_0}} =
     \zeta^{\pm 1} \psi(z,\pm g_0), \qquad \zeta^k = 1
     \eeq
then $\A (g^2)$ appears as the subalgebra of $\Zset_k$-invariant elements of
$\A (g_0^2)$ and the corresponding RCFT is named a $\Zset_k$-orbifold theory
of $\A (g_0^2).$

There are known local extensions of \Wc for arbitrary positive integer $c.$
These are: \\
\hspace*{12pt} (1) The $\Zset _2$-graded algebra $\A _c(1,-1)$ generated by $c$
pairs of free Fermi fields (see (\ref{1.15}), (\ref{1.16a})) or the associated
bosonic subalgebra generated by a pair of charge $\pm 2$ fields and by the
level 1 ${\mathrm{so}}(2c)$ currents.  \\
\hspace*{12pt} (2) Tensor products of $\widehat{\mathrm{su}}(c)_1$ theories
with any of the above $\A (g^2)$ algebras. \\
The (most degenerate) minimal \W models can be embedded in a number of rational
orbifold theories corresponding to finite subgroups of
$(\mathrm{SU}(c)\subset~)$ $\mathrm{SO}(2c)$ (cf. \cite{dv3}) that are
currently under study \cite{kt}. \\
\hfill \\
This paper was written during the stay of one of the authors, I.T., at the
Department of Mathematics, M.I.T., with the support of a Fulbright grant 19684.
I.T. thanks Victor Kac for hospitality and discussions. The work has been
supported in part by the Bulgarian Foundation for Scientific Research under
contract F-404.

\renewcommand{\refname}{ {\bf\normalsize\flushleft{References}} }
\begin{small}

\end{small}
\end{document}